\documentstyle[prl,aps,twocolumn]{revtex}
\large  

\input epsf
\begin{document}
\draft \twocolumn[\hsize\textwidth
\columnwidth\hsize\csname
@twocolumnfalse\endcsname

\title{Quantum revivals and carpets in some exactly solvable systems} 
\author{Will Loinaz$^{1}$\cite{wea} and T. J. Newman$^{1,2}$\cite{tea}} 
\address{$^{1}$Department of Physics, Virginia 
Polytechnic Institute and State University, Blacksburg 
VA 24061\\
$^{2}$Department of Physics, University of Virginia, McCormick Road,
Charlottesville VA 22903} 
\maketitle
\begin{abstract}
We consider the revival properties of quantum systems with an eigenspectrum
$E_{n} \propto n^{2}$, and compare them with the simplest member of this
class -- the infinite square well. In addition to having perfect revivals
at integer multiples of the revival time $t_{R}$, these systems all enjoy 
perfect fractional revivals at quarterly intervals of $t_{R}$. A closer
examination of the quantum evolution is performed for the 
P\"oschel-Teller and Rosen-Morse potentials,
and comparison is made with the infinite square well using quantum carpets.
\end{abstract}
\vspace{2mm} \pacs{PACS numbers: 42.50.Md, 03.65.Ge }]

\narrowtext

Over the past ten years or so there has been a growing interest in the 
quantum dynamics of simple systems, motivated in part by the richness of 
new phenomena such as revivals\cite{bkp} and quantum 
carpets\cite{grs,marz,roz}. Indeed, the
phenomenon of revivals is not merely a theoretical construct, but
has been observed in ion traps\cite{itr}, Rydberg atoms\cite{ra}, 
and semiconductor wells\cite{sw}; and has been used to differentiate
ionization pathways in potassium dimers\cite{rs}. Quantum revivals
are similar, but distinct from, quantum Poincar\'e recurrences,
which have been studied recently in the context of the kicked 
rotator\cite{qpr}. In the former, one is interested in the 
deterministic reconstruction
of the wave-function during its evolution inside a 
fixed potential; whilst in the
latter, interest is focused on the decay of the return probability
in ``mixed'' regions of phase space, as a measure of the quantum chaos 
in the (usually forced) system.
For the most part, analytic studies of revivals and carpets have 
concentrated on the infinite square well (ISW) potential, which is known to have 
perfect revivals {\it and} fractional revivals\cite{as}, and also a quantum carpet 
composed of rays (straight lines in the space-time plane)\cite{marz}.
These properties have been understood on the basis of the
quadratic dependence of the energy on quantum number, along
with the fact that the eigenfunctions are elementary trigonometric
functions. 

It is well understood that perfect revivals can only occur for systems
whose energy spectrum is purely quadratic in the quantum number\cite{bkp}. If the
dependence is purely linear (harmonic oscillator) the only time scale
is the classical period of oscillation, while for more complicated
energy spectra, revivals will be imperfect due to modulations from
the super-revival time scale. Although many studies have been devoted
to the simplest quantum system with a quadratic energy spectrum -- namely,
the ISW -- there has been less attention paid to the
host of other potentials which share this property (although see
Ref.\cite{evu} for a discussion of the autocorrelation function
for the Morse potential). It is guaranteed that these
systems will have perfect revivals, but what can one say about fractional
revivals, and the existence of quantum carpets ({\it i.e.} hidden structures 
in the space-time plot of the probability density)?

We shall begin with some very general remarks about fractional revivals.
Consider a system with a purely quadratic, nondegenerate energy 
spectrum $E_{n} = \alpha ^{2}n^{2}$, ($n=0,1,2,\cdots $), 
and with a potential $V(x)$ centered at $x=0$. We take the potential to be
an {\it even} function of $x$. In this case 
the eigenfunctions $\phi _{n}(x)$  will have a definite even or odd symmetry, 
alternating as the quantum number increases, the ground state naturally being 
even, since it has no node. So we have $\phi _{n}(-x) = (-1)^{n}\phi _{n}(x)$.
We prepare the wave function of the system in an initial state specified by
the energy eigenfunction expansion
\begin{equation}
\label{init}
\psi (x,0) = \sum \limits _{n} c_{n} \phi _{n}(x) \ .
\end{equation}
We restrict ourselves to contributions from bound states only.
The time-evolved wave function is given by
\begin{equation}
\label{time}
\psi (x,t) = \sum \limits _{n} c_{n} \phi _{n}(x) \exp [-iE_{n}t] \ ,
\end{equation}
where we have chosen units of $\hbar = 1$.
Given the quadratic dependence of the energy levels on $n$, it is easy
to see from Eq.(\ref{time}) that the wave function will be identical to
its initial state at integer multiples of the revival time 
$t_{R} \equiv 2\pi /\alpha ^{2}$.

Now consider the wave function at a time equal to one half of $t_{R}$.
One easily finds
\begin{equation}
\label{half1}
\psi (x,t_{R}/2) = \sum \limits _{n} c_{n} \phi _{n}(x) \exp [-i\pi n^{2}] \ .
\end{equation}
Given that $e^{-i\pi n^{2}}=(-1)^{n}$, we have
\begin{equation}
\label{half2}
\psi (x,t_{R}/2) = \sum \limits _{n \ {\rm even}} c_{n} \phi _{n}(x)
- \sum \limits _{n \ {\rm odd}} c_{n} \phi _{n}(x)  \ .
\end{equation}
Returning to the initial wave function, one may use the parity properties of 
the eigenstates to demonstrate that
\begin{eqnarray}
\label{parity}
\nonumber
\psi (x,0) & = & \sum \limits _{n \ {\rm even}} c_{n} \phi _{n}(x)
+ \sum \limits _{n \ {\rm odd}} c_{n} \phi _{n}(x) \ , \\
\psi (-x,0) & = & \sum \limits _{n \ {\rm even}} c_{n} \phi _{n}(x)
- \sum \limits _{n \ {\rm odd}} c_{n} \phi _{n}(x) \ .
\end{eqnarray}
Clearly, on comparing Eqs. (\ref{half2}) and (\ref{parity}) we have
the perfect fractional revival $\psi (x,t_{R}/2) = \psi (-x,0)$.
This result may appear to follow from the symmetry of the potential and
time reversal invariance; however, this is not the case  [cf. the discussion
following Eq.(\ref{halfsc})].

A less obvious result follows, however, when we study the wave function
at one quarter of the revival time. We have
\begin{equation}
\label{quart1}
\psi (x,t_{R}/4) = \sum \limits _{n} c_{n} \phi _{n}(x) \exp [-i\pi n^{2}/2] \ .
\end{equation}
Considering the phase for $n=0,1,2,3 \ {\rm mod}(4)$ one can easily establish
that
\begin{equation}
\label{quart2}
\psi (x,t_{R}/4) = \sum \limits _{n \ {\rm even}} c_{n} \phi _{n}(x)
-i \sum \limits _{n \ {\rm odd}} c_{n} \phi _{n}(x) \ .
\end{equation}
Solving the two expressions in Eq.(\ref{parity}) for the odd and even
sets of modes, we find the perfect fractional revival
\begin{equation}
\label{quart3}
\psi (x,t_{R}/4) = {(1-i) \over 2} \psi (x,0) + {(1+i)\over 2} \psi (-x,0) \ .
\end{equation}
In a similar manner one may show that
\begin{equation}
\label{quart4}
\psi (x,3t_{R}/4) = {(1+i) \over 2} \psi (x,0) + {(1-i)\over 2} \psi (-x,0) \ .
\end{equation}
Thus, a system with an even potential and a purely quadratic energy spectrum
supports perfect fractional revivals at quarters of the revival time.
Especially interesting are the fractional revivals at $t_{R}/4$ and $3t_{R}/4$
which for an initially localized wave function will consist of two
perfect, mirrored ``cat states''.
We have failed to find perfect fractional revivals at other fractions of 
the revival time for this general class of systems ({\it i.e.} utilizing
only parity properties of the eigenfunctions).

Let us now be more specific, and consider in turn two potentials of
the type considered above; namely symmetric cases of the P\"oschel-Teller (PT)
and Rosen-Morse (RM)
potentials\cite{ptrm,susy}. 
PT takes the form (with the ground state energy set at zero)
\begin{equation}
\label{scarf1}
V_{S1}(x) = -A^{2} + A (A-\alpha){\rm sec}^{2}(\alpha x) \ ,
\end{equation}
defined in the range $-\pi/2 \le \alpha x \le \pi/2$, and with energy spectrum 
\begin{equation}
\label{scarf1en}
E_{n} = (A+n\alpha)^{2} - A^{2} \ .
\end{equation}
In order to have perfect quarterly revivals, it is necessary to choose
$A=M\alpha$, with $M$ an integer.
Note that PT has an infinite number of bound states, and no
scattering states. The bound states may be expressed in terms of Gegenbauer
polynomials with argument ${\rm sin}(\alpha x)$. We shall restrict our attention
to $M=2$, in which case the energy eigenfunctions are sums of 
bilinear products of elementary trigonometric functions. 

RM takes the form
\begin{equation}
\label{scarf2}
V_{S2}(x) = A^{2} - A (A+\alpha ){\rm sech}^{2}(\alpha x) \ ,
\end{equation}
defined for $x$ on the entire real line, and with energy spectrum 
\begin{equation}
\label{scarf2en}
E_{n} = A^{2} - (A-n\alpha )^{2} \ .
\end{equation}
Again, to ensure perfect quarterly revivals, we choose $A=M\alpha $, with
$M$ a positive integer.
RM has only $M$ bound states, which may be expressed
in terms of the Gegenbauer polynomials with argument 
$i{\rm sinh} (\alpha x)$.

Although the energy spectra for these potentials are not purely
quadratic in $n$, it is a simple matter to redefine the quantum
number by shifting by $M$, in which case the perfect quarterly
revivals found above take the slightly modified form
\begin{eqnarray}
\label{halfsc}
\nonumber
\psi (x,t_{R}/4) & = & {1\over 2}(1-i \theta)\psi (x,0) + 
{1\over 2}(1+i\theta)\psi (-x,0) \ , \\
\psi (x,t_{R}/2) & = & \psi (-x,0) \ , \\
\nonumber
\psi (x,3t_{R}/4) & = & {1\over 2} (1+i\theta)\psi (x,0) + 
{1\over 2}(1-i\theta)\psi (-x,0) \ ,
\end{eqnarray}
where $\theta = (-1)^{M}$. If one chooses $A/\alpha $ to be
a half-integer, the revival at $t_{R}/2$ is {\it identical} to
the full revival, yet the perfect quarterly revivals are lost.

Much can be learnt about these systems by preparing an initial wave function
from a finite number of eigenstates, and then studying its evolution
using Eq.(\ref{time}). It is common practice\cite{bkp} to construct the initial wave
function using the first $N$ modes, with weights $c_{n}$ drawn from a 
Gaussian distribution centered at some reasonably energetic mode 
${\bar n}$ 
\begin{equation}
\label{Gauss}
|c_{n}|^{2} \sim \exp \left [ - {(n-{\bar n})^{2} \over 2 \sigma ^{2}} \right ] \ .
\end{equation} 
This simulates, for instance, a laser-prepared state in an ion trap. We have evolved 
the wave function in the PT and RM potentials
using these weights. [We choose equal phases for the $c_{n}$ in order
to create a well-localized initial wave packet. Choosing random phases
does not affect the revival structure, but tends to obscure the 
regularity of the quantum carpets.] Aside from confirming the perfect quarterly
revivals found above, we have found that PT has nearly perfect
fractional revival states at rational fractions of the revival time,
whereas there is little sign of such nearly perfect states for RM
(although they will appear for much higher energy wave packets\cite{bkp}).
In Figs.1 and 2 we illustrate this by showing the probability density $\rho $ at 
times $t=0, \ t_{R}/5, \ t_{R}/4$ and $t_{R}/3$ for PT and RM potentials
respectively. To test the robustness of the quarterly revivals we have also
evolved an excited wave packet in the RM potential, but away from the
rationality condition $A/\alpha = M$. 
We set $A/\alpha = M+r$, with $r \in (0,1/2)$. 
We find imperfect, yet ``smooth'' fractional revivals at 
$t_{R}/2$ for general values of $r$ (see Fig. 3a). 
However, the fractional revivals at $t_{R}/4$ are more
sensitive to $r$, and fade away for $r > 0.25$ (see Figs. 3b-3d). 

The almost perfect fractional revivals for PT may be
understood intuitively, since for a wave function constructed around
a moderately energetic state, the PT potential closely resembles
the ISW -- {\it i.e.} the harmonic structure of the
bottom of the well is barely resolved. [One may also make a more
quantitative argument by changing the basis of Eq.(\ref{time}) from
the PT eigenfunctions to the eigenfunctions of the ISW. 
One finds that the overlap integrals for large $n$ are
sharply peaked. Thus the dominant part of the wave function may be
expanded (with the weights $c_{n}$) in terms of the energy eigenstates of the
ISW, which, as mentioned before, has perfect fractional revivals
at all rational fractions of $t_{R}$.] Given the poor resolution of the structure of
the potential, one would also expect that the quantum carpets for
PT closely resemble those found for the ISW.
This is indeed the case, as shown on the left of Fig.4, where the characteristic
rays are clearly visible. 

\begin{figure}[htbp]
\epsfxsize=3.0in 
\hspace*{0.1cm}
\epsfbox{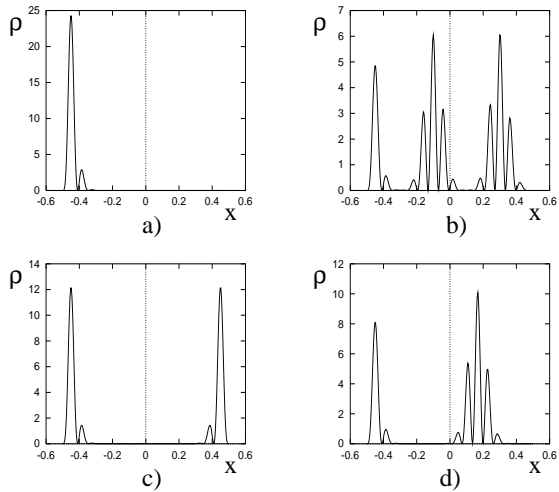}
\vspace*{0.5cm}
\caption{Probability density for the PT potential (with parameter
values $\alpha = \pi$, $M=2$, $N=30$, $\sigma = 3.0$, and ${\bar n}=15$) at times
a) $t=0$, b) $t=t_{R}/5$, c) $t=t_{R}/4$, and d) $t=t_{R}/3$.}
\end{figure}

As an alternative to preparing the initial wave function around some
energetic state, one can use a weighting that favors the lower-lying
states (using an exponentially decaying distribution for the $c_{n}$
for example). We have studied this case, and indeed the correspondence
with the ISW disappears, and the evolution of the
wave function is a slowly modulated classical oscillation (since
the well has a harmonic minimum). The almost perfect fractional
revivals are invisible, and the quantum carpet has no structure.
The perfect quarterly revivals may still be resolved.

\begin{figure}[htbp]
\epsfxsize=3.0in 
\hspace*{0.1cm}
\epsfbox{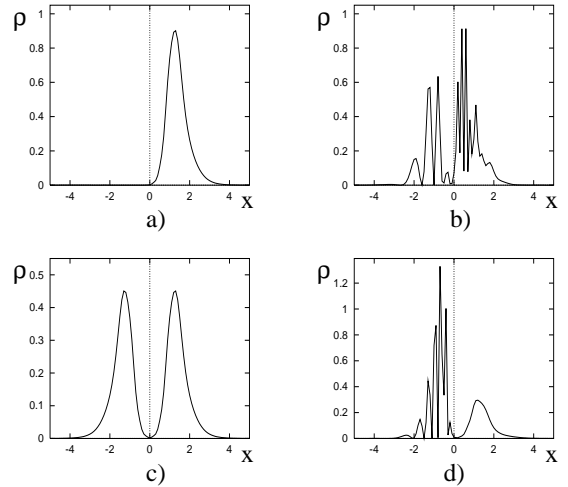}
\vspace*{0.5cm}
\caption{Probability density for the RM potential (with parameter
values $\alpha = 1.0$, $M=N=20$, $\sigma = 4.0$, and ${\bar n}=10$) at times
a) $t=0$, b) $t=t_{R}/5$, c) $t=t_{R}/4$, and d) $t=t_{R}/3$.}
\end{figure}

\begin{figure}[htbp]
\epsfxsize=3.0in 
\hspace*{0.1cm}
\epsfbox{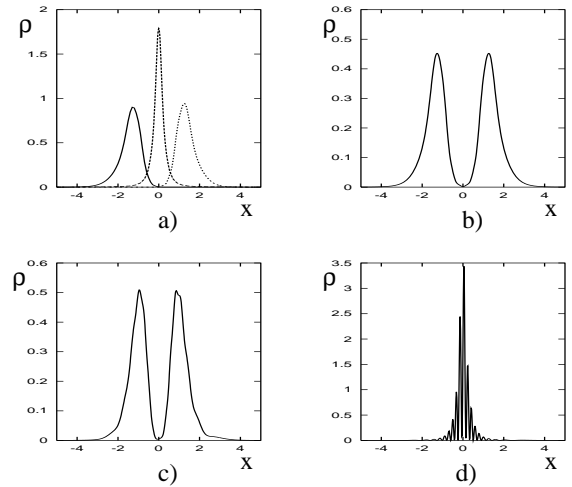}
\vspace*{0.5cm}
\caption{Probability density for the RM potential (with parameter
values $\alpha = 1.0$, $M=N=20$, $\sigma = 4.0$, and ${\bar n}=10$) for 
values of $A/\alpha = M+r$, with a) $r=0.0$ (leftmost), $0.25$ (centre), 
and $0.5$ (rightmost) at time $t_{R}/2$; b) $r=0.0$ at $t_{R}/4$; 
c) $r=0.25$ at $t=t_{R}/4$; and  d) $r=0.5$ at $t=t_{R}/4$.}
\end{figure}

As mentioned above, the RM potential shows little sign of
fractional revivals, apart from the perfect quarterly revivals.
However, the quantum carpet for this potential reveals considerable
structure, as shown on the right of Fig.4; the rays of PT are replaced by 
a complicated structure of what appear to be non-linear ``world lines''. A
magnified view of the first perfect quarterly revival is shown in Fig.5.

\begin{figure}[htbp]
\epsfxsize=3.0in 
\vspace*{-0.2cm}
\hspace*{0.2cm}
\epsfbox{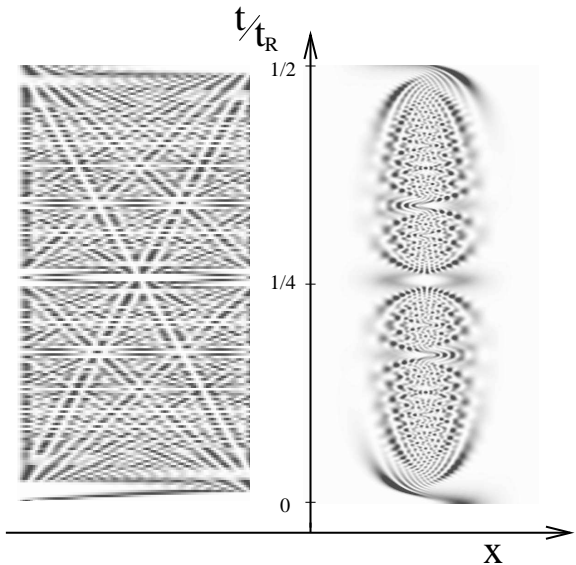}
\vspace*{0.3cm}
\caption{The quantum carpets ({\it i.e.} space-time contour plots of the probability 
density) for PT (left) and RM (right), with parameter values
as before, shown for $t \in (0,t_{R}/2)$. The darker regions indicate higher
probability density.}
\end{figure} 

\begin{figure}[htbp]
\epsfxsize=3.0in
\vspace*{0.1cm}
\hspace*{0.2cm}
\epsfbox{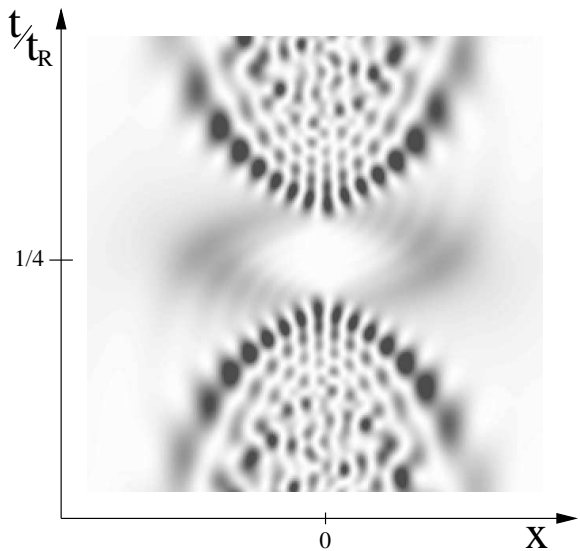}
\vspace*{0.2cm}
\caption{Detail of the quantum carpet for the RM potential around
$t=t_{R}/4$, showing the emergence and subsequent collapse 
of two perfect cat states.}
\end{figure} 

We have yet to attain a physical understanding of the quantum
carpet for RM. Approaches which work well for the ISW
\cite{marz} are less useful here due to the complicated nature of the eigenfunctions.
It is interesting to note that the structures visible in the ISW/PT
quantum carpet ({\it i.e.} rays) can be interpreted as a superposition of 
coherent wave packets (CWP) which follow classical trajectories (with a 
discrete spectrum of initial velocities selected to ensure a perfect revival
at $t=t_{R}$). Whether these CWP can be identified with
the coherent states\cite{ks} corresponding to the ISW is unclear. The
apparent world-lines of the RM quantum carpet might also be interpreted
in this way, although the CWP no longer follow classical trajectories. This
is clear from the manner in which the world lines proceed through
the minimum of the potential at $t=t_{R}/4$. 


A computational application of our results is the testing of numerical algorithms
designed to integrate forward the time-dependent Schr\"odinger equation\cite{hans}. 
These algorithms do not generally rely upon energy eigenfunction expansions, and
testing them against exact results (for general initial conditions, and
non-trivial potentials) is difficult 
due to the scarcity of such results in quantum dynamics. An algorithm which integrates
the wave function forward in time in the PT or RM potentials, and 
successfully generates ({\it i.e.} recovers with good precision) perfect quarterly 
revivals can be trusted in other applications. A positive feature of this test 
is that one can implement
it for {\it any} initial condition (strictly true only for PT for which
the bound states form a complete set). 
It is interesting to note that
algorithms which use discrete Fourier modes for free-particle propagation will fail
to capture perfect revivals, since they are based
on a Hamiltonian with a discrete lattice Laplacian, 
thus replacing the pure $k^{2}$ 
spectrum by $2(1 - {\rm cos}k )$,
although they will find increasingly good revival
structures as the number of Fourier modes is increased.

In conclusion, we have studied the quantum revivals and carpets for
systems with a quadratic energy spectrum and an even potential.
We have proven that {\it all} such systems have perfect quarterly revivals,
in addition to perfect complete revivals. We have studied the
time evolution of two such systems -- the PT and RM potentials
-- more closely. From a moderately energetic initial distribution of modes,
the evolution of the wave function in PT (being 
defined in a finite region of space) has many similarities to that
of the ISW, with almost perfect fractional revivals
at rational fractions of $t_{R}$, and a quantum carpet with
characteristic rays. This similarity disappears continuously as one
decreases the mean energy of the wave function, thus allowing better
resolution of the harmonic minimum of the well. The evolution of
the wave function in RM shows little fractional revival structure,
apart from the perfect quarterly revivals. Its quantum carpet, although
devoid of rays, displays a dazzling pattern, the understanding of which
is currently being pursued.

This study has shown that
many of the rich dynamical properties of the ISW
are fairly generic, thus increasing their experimental relevance. 
Indeed, the PT and RM
potentials capture features of real quantum systems, which are
missing in the ISW. Namely, a spatially varying
potential energy, with a harmonic minimum; and in the case of
RM, a finite number of bound states (see also the discussion
of super-revivals in the finite square well\cite{va}). The robustness
of the perfect quarterly revivals may well be of interest to 
experimentalists seeking to create perfect cat states from
a localized wave packet. Indeed, this is the initial entangled
state required for quantum communication\cite{qcomm} (although 
``entanglement'' usually refers to two or more degrees of freedom), and which 
is generally
created using more complicated laser interferometry. (Fabrication of an 
approximate RM potential may well be realisable using semiconductor 
quantum well technology.) 

Aside from the PT and RM 
potentials, there are other potentials which are isospectral to the
ISW (and which may be generated using the Darboux
transformation\cite{susy}). A study of their revival properties may well prove
worthwhile.
As a final remark, it is noteworthy that the PT potential (with $M=2$) and the 
ISW are supersymmetric partner potentials\cite{susy}, and therefore
share the same energy spectrum (bar the lowest state). Whether,
due to supersymmetry, these systems share
other {\it dynamical} equivalents, aside from perfect quarterly revivals, 
is an interesting open question.

\vspace{0.2cm}

T. J. N. thanks Robert R. Jones, Jr for several illuminating discussions.
W. L. gratefully acknowledges financial support from the Department of Energy.
T. J. N. gratefully acknowledges financial support from the Division of Materials
Research of the National Science Foundation.

\end{document}